\documentclass[printer]{aa}

\bibpunct{(}{)}{;}{a}{}{,}

\usepackage{lscape}
\usepackage{graphicx}
\usepackage[varg]{txfonts}
\newcommand{\hii}{H\textsc{ii}}
\def\ks{km s$^{-1}$}

\def\m{$^\prime$}
\def\s{$^{\prime\prime}$}

\def\cm3{cm$^{-3}$}

\def\2{$^{12}$CO}
\def\3{$^{13}$CO}
\def\8{C$^{18}$O}

\def\cm2{cm$^{-2}$}

\def\x{X$^{13/18}$}
\def\g29{G29.96$-$0.02}

\begin{document}

\title{Mapping the $^{13}$CO/C$^{18}$O abundance ratio in the massive star forming region G29.96$-$0.02}
\author{S. Paron \inst{1,2}
\and M. B. Areal \inst{1}
\and M. E. Ortega \inst{1}
}

\institute{CONICET-Universidad de Buenos Aires. Instituto de Astronom\'{\i}a y F\'{\i}sica del Espacio
             CC 67, Suc. 28, 1428 Buenos Aires, Argentina 
\and Universidad de Buenos Aires. Facultad de Arquitectura, Dise\~{n}o y Urbanismo. Buenos Aires, Argentina
}

\offprints{S. Paron}

   \date{Received <date>; Accepted <date>}

\abstract{}
{Estimating molecular abundances ratios from the direct measurement of the emission of the molecules towards 
a variety of interstellar environments is indeed very useful to advance in our understanding
of the chemical evolution of the Galaxy, and hence of the physical processes related to the chemistry. It is necessary to increase
the sample of molecular clouds, located at different distances, in which the behaviour of molecular abundance ratios, such as 
the \3/\8 ratio, is studied in detail.}
{We selected the well-studied high-mass star-forming region \g29, located at a distance of about 6.2 kpc, which is an ideal laboratory 
to perform this kind of studies. To study the \3/\8 abundance ratio (\x) towards this region it was used \2 J=3--2 data obtained from 
the CO High-Resolution Survey, \3 and \8 J=3--2 data from the \3/\8 (J=3--2) Heterodyne Inner Milky Way Plane Survey, and \3 and \8 
J=2--1 data retrieved from the CDS database which were observed with the IRAM 30 m telescope. The distribution of column densities and 
\x~throughout the extension of the analyzed molecular cloud was studied based on LTE and non-LTE methods.}
{Values of \x~between 1.5 to 10.5, with an average of about 5, were found across the studied region, showing that,
besides the dependency between \x~and the galactocentric distance, the local physical conditions may strongly affect this 
abundance ratio. We found that correlating the \x~map with
the location of the ionized gas and dark clouds allows us to suggest in which regions the far-UV radiation stalls in dense gaseous components,
and in which ones it escapes and selectively photodissociates the \8 isotope.
The non-LTE analysis shows that the molecular gas has very different physical conditions, not only spatially across the cloud, but 
also along the line of sight. This kind of studies may represent a tool to indirectly estimate (from molecular lines observations) 
the degree of photodissociation in molecular clouds, which is indeed useful to study the chemistry in the interstellar medium.

}
{}

\titlerunning{Mapping the $^{13}$CO/C$^{18}$O abundance ratio in G29.96$-$0.02}
\authorrunning{S. Paron et al.}

\keywords{ISM: abundances -- ISM: molecules -- Galaxy: abundances --  {\it (ISM:)} H\textsc{ii} regions --  Stars: formation   }

\maketitle

\section{Introduction}

Given the current availability of large molecular lines surveys it is now possible to estimate molecular abundances ratios
from the direct measurement of the emission of the molecules themselves towards a variety of interstellar environments. 
This avoids the use of indirect estimations from known elemental abundances, such as the case of the \3/\8 abundance ratio (\x), 
that in general it is estimated from a double ratio between the
$^{12}$C/$^{13}$C and $^{16}$O/$^{18}$O ratios given by \citet{wilson94}. In a previous paper \citep{areal18}, 
using \2, \3, and \8 J=3--2 data we studied the \x~towards a large sample of young stellar objects (YSOs) 
and \hii~regions distributed along the first Galactic quadrant, finding ratios systematically lower than the predicted from the mentioned 
elemental abundance relations. 
We also found that the ratios depend not only on the distance to the Galactic center, as shown by \citet{wilson94}, but also on 
the type of source or region observed. YSOs tend to have smaller \x~than \hii~regions, which proves 
the selective far-UV photodissociation of the \8. 

These results were in agreement with some previous works that studied the relation between
\x~and the far-UV radiation towards different regions in nearby molecular clouds with different physical conditions. For instance, 
\citet{lada94} found in IC 5146, a low-mass star-forming region, a \x~value considerably greater than the expected solar value
in the outer parts of the cloud. The same was found by \citet{shima14} in most of the studied regions towards Orion-A, which are affected by
OB stars. \citet{kong15} observed a gradient in \x~with the decreasing of the $A_{\rm V}$ in the southeast of the California molecular cloud,
while the same was observed in L 1551, an isolated nearby star forming region that does not contain OB stars and thus it is influenced only by
the far-UV from the interstellar radiation \citep{lin16}.

This kind of studies towards molecular clouds harboring (or not) different objects, such as YSOs, OB stars, \hii~regions, bubbles, etc,
are certainly useful to advance in our understanding 
of the chemical evolution of the Galaxy, and hence of the physical processes related to the chemistry. Thus, it is necessary to increase
the sample of molecular clouds, located at different distances, in which the behaviour of the \x~value is studied in detail throughout the
extension of the cloud. We selected the well-studied high-mass star-forming region \g29, a quite unique region, to perform that.

\g29 has an angular size of about 12\m, it is located at a distance of 6.2 kpc \citep{russ11}, and it harbors 
different objects: a cometary 
ultracompact \hii~region \citep{cesaroni94} ionized by a known O star \citep{watson97} which is associated with a hot molecular 
core \citep{wood89,cesaroni94,olmi03}; an infrared dark cloud with multiple massive low-temperature cores \citep{pillai}; several 
radio continuum sources (NVSS sources) and many YSOs in different evolutionary stages \citep{beltran13}. The molecular gas 
related to this region was studied by \citet{carl13} using \3 and \8 J=2--1 data obtained with the IRAM 30 m telescope.
\g29 also presents diffuse X-rays emission \citep{towns14},
which as the authors point out it is quite interesting because may indicate that massive star winds likely perforate their natal clouds
and contribute to the hot of the interstellar medium. 
As seen, this region was multiwavelength observed and analyzed, resulting in 
a quite complete characterization of its physical conditions. Thus, \g29 is an ideal laboratory to study the relation between \x~and
the molecular gas conditions throughout the cloud.

Finally, it is important to mention that an abundance ratio study can be done by assuming local thermodynamic equilibrium (LTE) from
the emission of an unique transition from several isotopes (the simplest way), or by using non-LTE procedures. The last one is difficult 
to perform in large regions because observations/surveys of two or more transitions of same molecules/isotopes are needed. In this work we 
present results from both methods.

\begin{figure*}[tt]
\centering
\includegraphics[width=15cm]{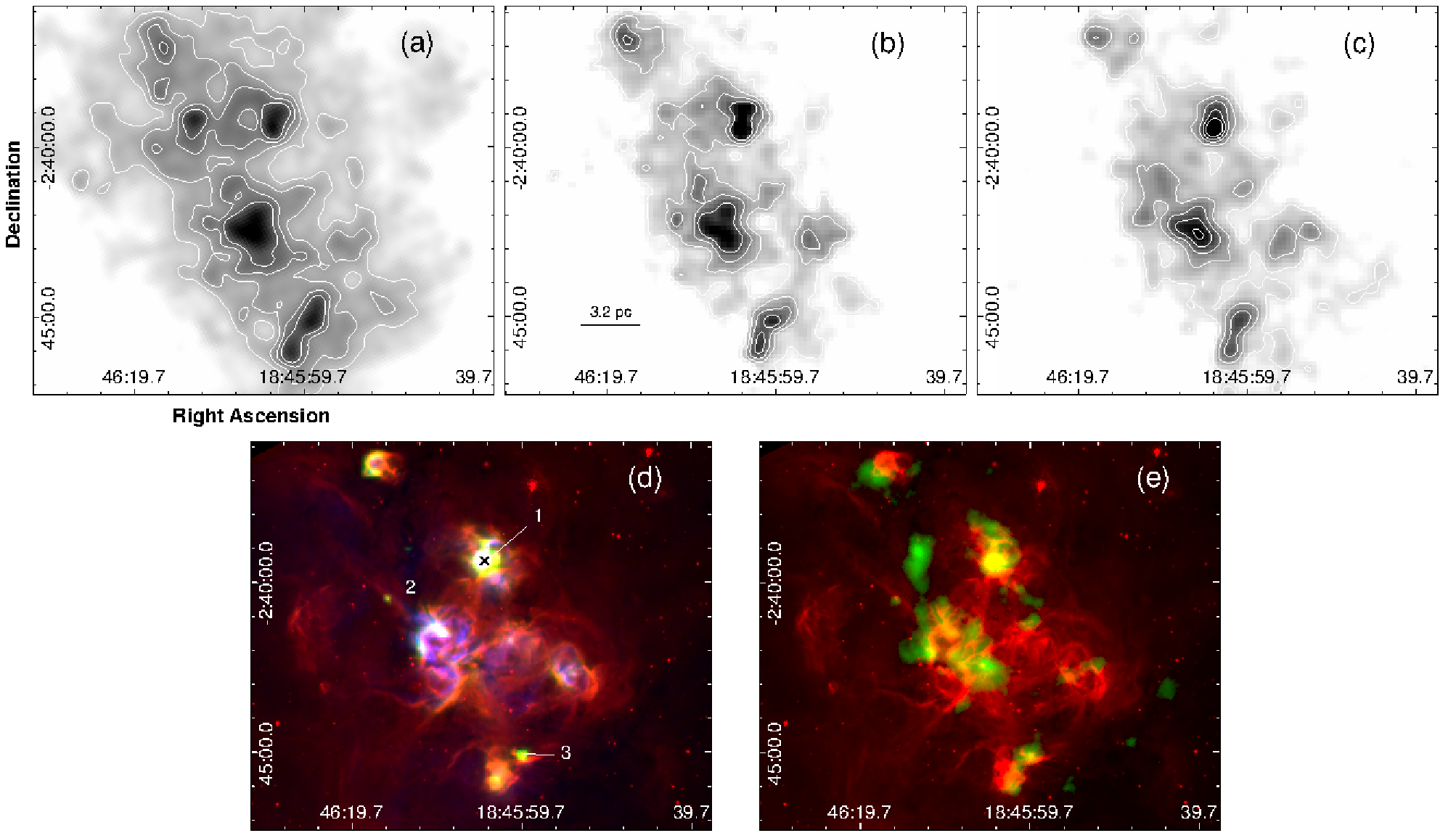}
\caption{(a) \2 J=3--2 line integrated between 80 and 120 \ks, the contour levels
are: 130, 160, 196, 230, and 262 K \ks. (b) \3 J=3--2 line integrated between 90 and 110 \ks, the contour levels
are: 27, 41, 55, and 72  K \ks. (c) \8 J=3--2 line integrated between 90 and 110 \ks~with contour levels of 5, 11, 16, 22 and 27 K \ks.
(d) Three-colour image towards \g29~displaying the
{\it Spitzer}-IRAC 8 $\mu$m emission in red, the {\it Herschel}-PACS 70 $\mu$m emission in green, and the radio continuum
emission at 20 cm as extracted from the MAGPIS in blue. The numbers indicate the position of some features described in the text.
The black $\times$ indicates the position of the known O-type star \citep{watson97}. 
(e) Two-colour image of \g29~region showing {\it Spitzer}-IRAC 8 $\mu$m emission in red   
and the emission at 850 $\mu$m obtained from SCUBA in green.}
\label{g29present}
\end{figure*}

\section{Data}

\subsection{J=2--1 line}

We used \3 and \8 J=2--1 data observed with the IRAM 30 m telescope towards \g29~by \citet{carl13}. The data cubes 
were obtained from the CDS\footnote{http://cdsarc.u-strasbg.fr/viz-bin/qcat?J/A+A/560/A24} and have angular and spectral resolutions of
11\farcs7 and 0.15 \ks, respectively. The intensity of the data are on main beam brightness 
temperature (T$_{\rm mb}$) scale. As indicated by \citet{carl13} the typical rms noise level of the spectra, in units
of T$_{\rm mb}$, is about 1 K for both isotopes.

\subsection{J=3--2 line}

The \2, \3, and \8 J=3--2 data were extracted from two public databases performed with the 15 m James Clerck Maxwell 
Telescope (JCMT) in Hawaii.
The \2 J=3--2 data were obtained form the CO High-Resolution Survey (COHRS) with an angular and spectral resolution of
14\s~and 1 \ks~\citep{dempsey13}. The data of the other isotopes were obtained from the
\3/\8 (J=3--2) Heterodyne Inner Milky Way Plane Survey (CHIMPS), which have an angular and spectral resolution of
15\s~and 0.5 \ks~\citep{rigby16}. The intensities of both set of data are on the $T_{A}^{*}$ scale, and it 
was used the mean detector efficiency $\eta_{\rm mb} = 0.61$ for the \2, and $\eta_{\rm mb} = 0.72$ for the \3 and \8 
to convert $T_{A}^{*}$ to main beam brightness temperature (T$_{\rm mb} = T_{A}^{*}/\eta_{\rm mb}$)  \citep{buckle09}.
The typical rms noise levels of the spectra, in units of $T_{A}^{*}$, are: 0.25, 0.35, and 0.40 K for
the \2, \3, and \8, respectively.

The data were visualized and analyzed with the Graphical Astronomy and Image Analysis Tool (GAIA)\footnote{GAIA is a derivative of 
the Skycat catalogue and image display tool, developed as part of the VLT project at ESO. Skycat and GAIA are free software under 
the terms of the GNU copyright.} and with tools from the Starlink software package \citep{currie14}.

\section{Presentation of \g29}

Figure\,\ref{g29present} presents the massive star-forming region \g29~as seen in different wavelengths. 
The molecular gas distribution related to \g29~is presented in panels (a), (b), and (c) where it is shown the integrated \2, \3, and \8 J=3--2
emission, respectively. Figure\,\ref{g29present} (d) is a three-colour image displaying the
{\it Spitzer}-IRAC 8 $\mu$m emission in red, the {\it Herschel}-PACS 70 $\mu$m emission in green, and in blue, the radio continuum
emission at 20 cm as extracted from the New GPS of the Multi-Array Galactic Plane Imaging Survey (MAGPIS).
Figure\,\ref{g29present} (e) displays in a two-colour image the {\it Spitzer}-IRAC 8 $\mu$m  and the SCUBA 
850 $\mu$m emission in red and green, respectively. The emissions presented in panels (d) and (e) were selected
in order to appreciate the borders of the photodissociation regions (PDRs) (displayed in the 8 $\mu$m emission), the 
distribution of the ionized gas (shown with the 20 cm emission) and the warm dust (70 $\mu$m emission) which are 
useful to indirect and roughly discern features and its degree of ionization and warming by the UV radiation, together 
with the distribution of cold dust (displayed in the 850 $\mu$m emission). Numbers included in  Figure\,\ref{g29present} (d)
show the positions of some interesting objects for reference: (1) an UC \hii~region associated with a hot molecular cloud \citep{beuther07},
and the black $\times$ indicates the location of its exciting O-type star \citep{watson97},
(2) an infrared dust cloud (better traced in the 850 $\mu$m emission in panel (e), see \citealt{pillai}), and (3) a massive YSO with 
molecular outflows (work in preparation). The linear size included in Fig.\,\ref{g29present} (b) and in subsequent figures
along this work is derived by assuming the distance of 6.2 kpc to the analyzed molecular cloud.

\begin{figure*}[tt!]
\centering
\includegraphics[width=19cm]{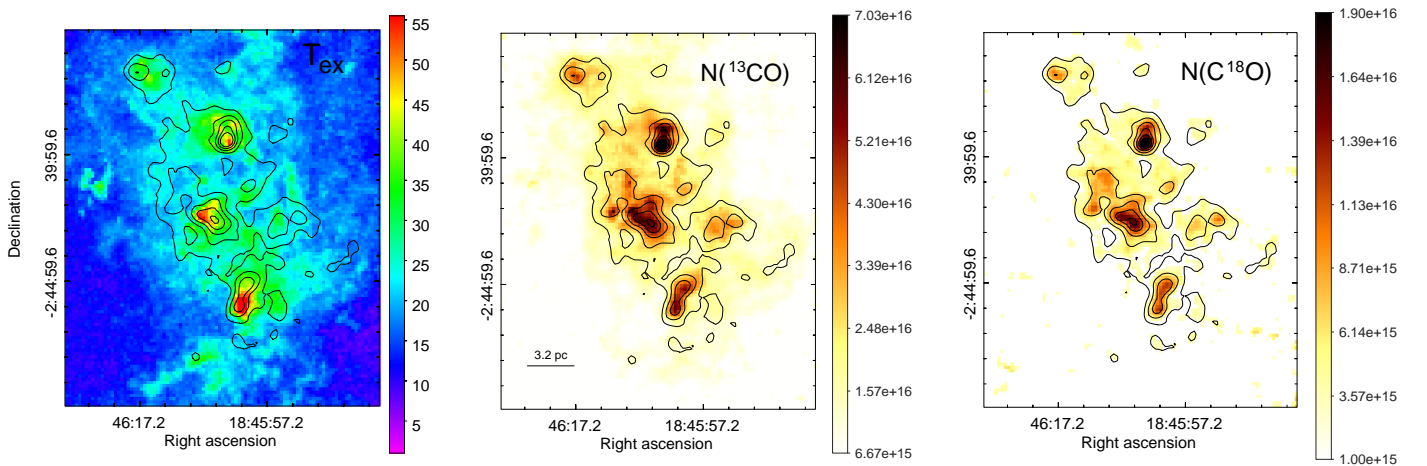}
\caption{Maps showing the results obtained from the \2, \3, and \8 J=3--2 emission under the LTE assumption. 
Left panel: excitation temperature
(T$_{ex}$), the colourbar is in units of K. Middle and right panels: \3 and \8 column densities, respectively. The colourbars are in
units of cm$^{-2}$. Contours of the integrated \8 J=3--2 emission are included for reference (same contours as presented 
in Fig.\,\ref{g29present} (c)).} 
\label{resultsLTE}
\end{figure*}

\section{Methodology and results}

We used two methods to estimate \3 and \8 column densities and \x (which is the ratio between them)
throughout the molecular cloud depicted by the CO molecular emission (see Fig.\,\ref{g29present} (a), (b), and
(c)). In the following subsections we describe both with their results.

\subsection{LTE analysis}

Assuming LTE we estimated the column densities of \3 and \8 pixel-by-pixel from the J=3--2 emission. The optical depths
($\tau_{\rm ^{13}CO}$ and $\tau_{\rm C^{18}O}$) and column densities
(N(\3) and N(\8)), assuming a beam filling factor of 1, were derived using the following equations:
\begin{equation}
  \tau_{\rm ^{13}CO} = - ln\left(1 - \frac{T_{mb}({\rm ^{13}CO})}{15.87\left[\frac{1}{e^{15.87/T_{ex}}-1} - 0.0028\right]}\right)    
\label{tau13}
\end{equation}

\begin{equation}
{\rm N(^{13}CO)} = 8.28 \times 10^{13}~e^{\frac{15.85}{T_{ex}}}\frac{T_{ex} + 0.88}{1 - e^{\frac{-15.87}{T_{ex}}}} 
\int{\tau_{\rm ^{13}CO}{\rm dv}}
\label{N13}
\end{equation}
with 
\begin{equation}
\int{\tau_{\rm ^{13}CO}{\rm dv}} = \frac{1}{J(T_{ex}) - 0.044} \frac{\tau_{\rm ^{13}CO}}{1-e^{-\tau_{\rm ^{13}CO}}}
\int{T_{\rm mb}({\rm ^{13}CO}){\rm dv}}
\label{integ13}
\end{equation}
\begin{equation}
  \tau_{\rm C^{18}O} = - ln\left(1 - \frac{T_{mb}({\rm C^{18}O})}{15.81\left[\frac{1}{e^{15.81/T_{ex}}-1} - 0.0028\right]}\right)    
\label{tau18}
\end{equation}
\begin{equation}
{\rm N(C^{18}O)} = 8.26 \times 10^{13}~e^{\frac{15.80}{T_{ex}}}\frac{T_{ex} + 0.88}{1 - e^{\frac{-15.81}{T_{ex}}}} 
\int{\tau_{\rm C^{18}O}{\rm dv}}
\label{N18}
\end{equation}
with 
\begin{equation}
\int{\tau_{\rm C^{18}O}{\rm dv}} = \frac{1}{J(T_{ex}) - 0.045} \frac{\tau_{\rm C^{18}O}}{1-e^{-\tau_{\rm C^{18}O}}}
\int{T_{\rm mb}({\rm C^{18}O}){\rm dv}}
\label{integ18}
\end{equation}
The $J(T_{ex})$ parameter is $\frac{15.87}{exp(\frac{15.87}{T_{ex}}) - 1}$ in the case of Eq.\,(\ref{integ13}) and 
$\frac{15.81}{exp(\frac{15.81}{T_{ex}}) - 1}$ in Eq.\,(\ref{integ18}).
In all equations $T_{\rm mb}$ is the peak main brightness temperature 
and $T_{ex}$ the excitation temperature. Assuming that the \2 J=3--2 emission is optical thick the $T_{ex}$ was derived
from:
\begin{equation}
T_{ex} = \frac{16.6}{{\rm ln}[1 + 16.6 / (T_{\rm peak}(^{12}{\rm CO}) + 0.036)]}
\label{tex}
\end{equation}

\begin{figure}[h]
\centering
\includegraphics[width=9.2cm]{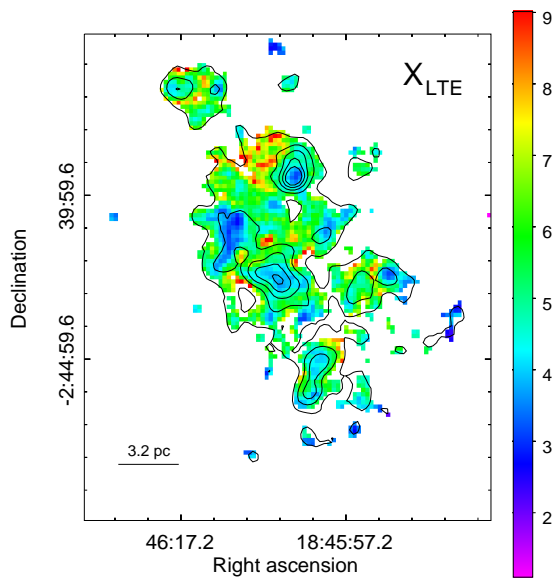}
\caption{Map of the \x~in the \g29~region obtained from the \2, \3, and \8 J=3--2 line under the LTE assumption.
Contours of the integrated \8 J=3--2 emission are included for reference.}
\label{Xlte}
\end{figure}

The \3 and \8 J=3--2 emissions are optically thin throughout the analyzed molecular cloud. The average value 
of $\tau_{\rm ^{13}CO}$ obtained from the whole region of the \3 emission is 0.44, with a minimum of 0.17, and a maximum of 1.1.  
In the case of $\tau_{\rm C^{18}O}$ it was obtained an average of 0.13 with a minimum of 0.04, and a maximum of 0.45.
Figure\,\ref{resultsLTE} presents the results obtained from the above expressions, displaying
maps of T$_{ex}$ (left panel), N(\3) and N(\8) (middle and right panels), and for reference contours of the
integrated \2 J=3--2 emission are included. Figure\,\ref{Xlte} displays a map of \x~towards \g29. Given that
the \8 line is weaker than the \3 line, and even undetectable in some regions of the cloud where \3 is detected, 
it is important to be cautious with not real low values in N(\8) which will yield artificial high values in \x. 
Thus, only values above $5\sigma$ from the \8 integrated map were considered, which implies  
values above $\sim 3\times10^{15}$ cm$^{-2}$ in the N(\8) map.

\begin{figure}[h]
\centering
\includegraphics[width=9cm]{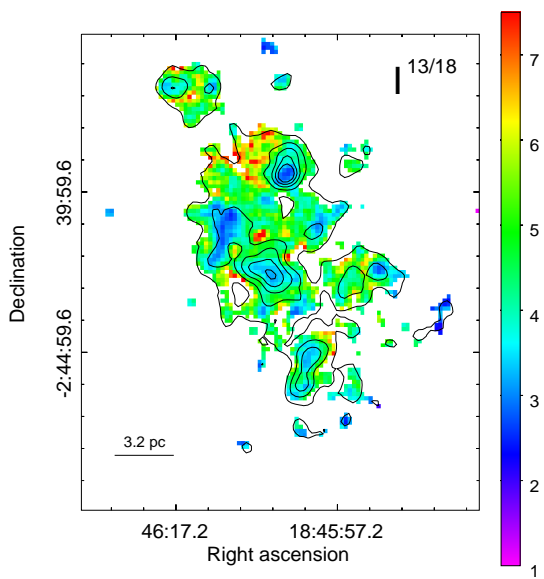}
\caption{Map of the integrated line ratio I$^{13/18}$ in the \g29~region.
Contours of the integrated \8 J=3--2 emission are included for reference.}
\label{mapIr}
\end{figure}

\begin{figure}[h]
\centering
\includegraphics[width=9cm]{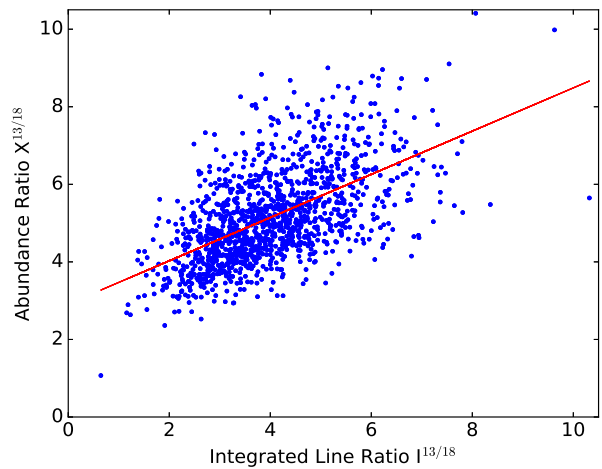}
\caption{Abundance ratio X$^{13/18}$ vs. integrated line ratio I$^{13/18}$ obtained from all
pixels between Figs.\,\ref{Xlte} and \ref{mapIr}. The red line is the result from a linear fitting.}
\label{plotIvX}
\end{figure}

\begin{figure}[h]
\centering
\includegraphics[width=9cm]{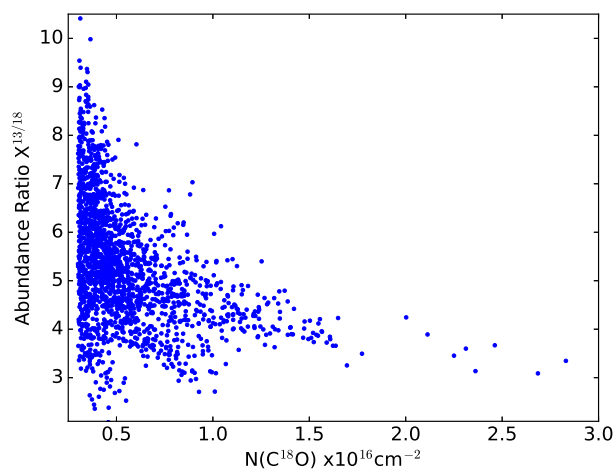}
\caption{Abundance ratio X$^{13/18}$ vs. the \8~column density obtained from all
pixels of Fig.\,\ref{Xlte}.}
\label{plotXvN18}
\end{figure}

Figure\,\ref{mapIr} displays a map of the integrated line ratio I$^{13/18}$ ($\int{T_{mb}^{13}~\rm{dv}}/ \int{T_{mb}^{18}~\rm{dv}}$)
and Fig.\,\ref{plotIvX} shows the abundance ratio X$^{13/18}$ vs. the integrated line ratio I$^{13/18}$ obtained from all
pixels of Figs.\,\ref{Xlte} and \ref{mapIr}. From a linear fitting to the data of the  X$^{13/18}$ vs. I$^{13/18}$ plot 
we obtain: $A = 0.58 \pm 0.08$ and $B = 2.80 \pm 0.02$ ($A x + B$) with R$^{2}=0.35$ (red line in Fig.\,\ref{plotIvX}).
Finally, Fig.\,\ref{plotXvN18} presents the relation between the abundance ratio and the \8 column density.

\subsection{Non-LTE analysis}

To estimate the column densities of \3 and \8  we used their J=2--1
and J=3--2 transitions and the RADEX code\footnote{RADEX is a statistical equilibrium radiative transfer code,
available as part of the Leiden Atomic and Molecular Database (http://www.strw.leidenuniv.nl/moldata/).} \citep{vander06}.
The \3 and \8 J=2--1 data cubes were degraded to the angular and spectral resolution of the \3 and \8 J=3--2 data. Then,
the spatial component of all cubes were rebinned to a $60\times73$ grid with pixels of about 11\s~in size.

\begin{figure}[h]
\centering
\includegraphics[width=8.8cm]{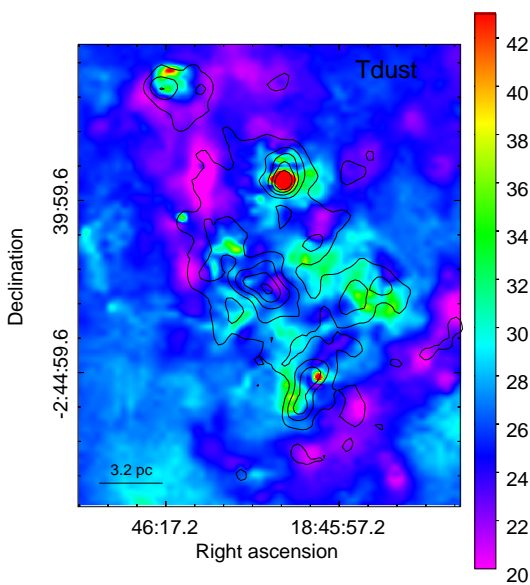}
\caption{Dust temperature derived from three {\it Herschel} ~bands (160, 250, and 350 $\mu$m) and the SCUBA band at 850 $\mu$m.
Contours of the integrated \8 J=3--2 emission are included for reference. The colour
bar is in units of K.}
\label{dust}
\end{figure}

\begin{figure}[h]
\centering
\includegraphics[width=7.9cm]{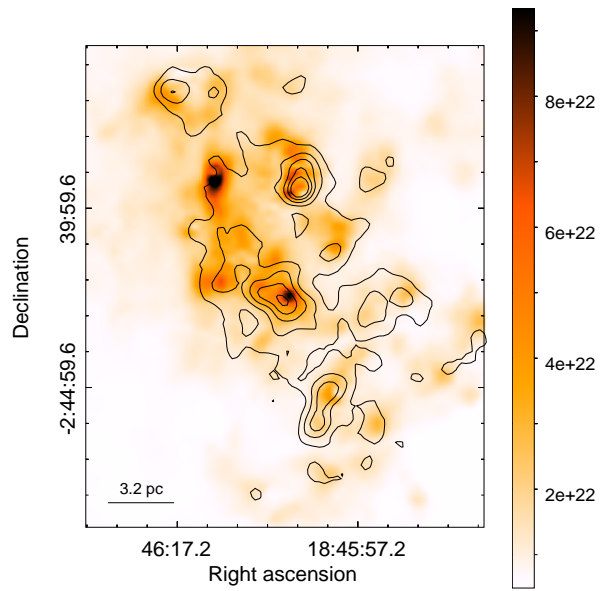}
\caption{H$_{2}$ column density derived from three {\it Herschel} ~bands (160, 250, and 350 $\mu$m) and the SCUBA band at 850 $\mu$m.
The colourbar is in units of cm$^{-2}$. Contours of the integrated \8 J=3--2 emission are included for reference.}
\label{nh2}
\end{figure}

The inputs of RADEX are: kinetic temperature (T$_{\rm K}$), line velocity width at FWHM ($\Delta$v), and
line peak temperature (T$_{\rm peak}$), and then the code fits the column and volume molecular densities (N and n$_{\rm H_{2}}$).
For the T$_{\rm K}$ we assumed that the gas is coupled to
the dust (i.e. T$_{\rm K} =$ T$_{\rm dust}$), and we performed a dust temperature map
obtained from a spectral energy distribution (SED) using data from {\it Herschel}-PACS at 160 $\mu$m, {\it Herschel}-SPIRE at 250 and 
350 $\mu$m, and SCUBA at 850 $\mu$m. The images at 160, 250 and 850 $\mu$m were convolved to the angular resolution of the {\it Herschel}-SPIRE
350 $\mu$m map (which has the coarsest angular resolution among all used bands, this is 24\s) and rebinned to the same pixel size of the molecular 
lines data cubes. Then, fluxes at the four wavelengths were fitted pixel-by-pixel with a modified black body curve:
\begin{equation}
S_{\nu} = \mu m_{H} {\rm N(H_{2})} k_{\nu{_0}} (\nu/\nu_{0})^{\beta} B_{\nu}({\rm T_{dust}})
\label{sed}
\end{equation}
\noindent where $S_{\nu}$ is the flux at the frequency $\nu$, $m_{H}$ the hydrogen atom mass, $\mu = 2.8$ by adopting a relative
helium abundance of 25\%, N(H$_{2}$) the H$_{2}$ column density, $B_{\nu}$ is the Plank function, 
$k_{\nu} =  k_{\nu{_0}} (\nu/\nu_{0})^{\beta}$ the dust opacity,
where we adopt $\beta = 2$, and for a frequency $\nu_{0} = 1$ THz and a gas-to-dust ratio of 100, $k_{\nu{_0}} = 0.1$ cm$^{2}$ g$^{-1}$.
The obtained dust temperature map (Figure\,\ref{dust}) shows almost the same results as presented in \citet{carl13}, who
followed the same procedure, but without the 850 $\mu$m data. We included this wavelength in our SED with the aim
of obtaining more accurate results towards the coldest regions. As an additional result obtained from this SED, we present in 
Figure\,\ref{nh2} the H$_{2}$ column density map towards the region.

To obtain $\Delta$v and T$_{\rm peak}$ pixel-by-pixel from the data cubes we performed maps of the second moment and the emission peak, 
respectively for the \3 and \8 J=2--1 and J=3--2 lines. Assuming Gaussian line profiles, the second
moment, which is the line velocity dispersion, was multiplied by a factor of 2.355 to obtain the FWHM $\Delta$v.
The molecular data were convolved to the same angular resolution of the T$_{dust}$ map (i.e. 24\s).
Even though \citet{carl13} pointed out that the velocity is rather homogeneous across this cloud (W43-South in their paper),
from our inspection of the data cubes we found that the central velocity and the line shape of the spectra change 
across the cloud. Thus, we carefully selected four regions in which the spectra allowed us to obtain reliable FWHM $\Delta$v maps. 
This is because if we integrate along a velocity range with low S/N ratio, the second moment can yield very high unrealistic
values. The selected velocity ranges for each line in each region are presented in Table\,\ref{ranges}.

\begin{figure}[h]
\centering
\includegraphics[width=9.3cm]{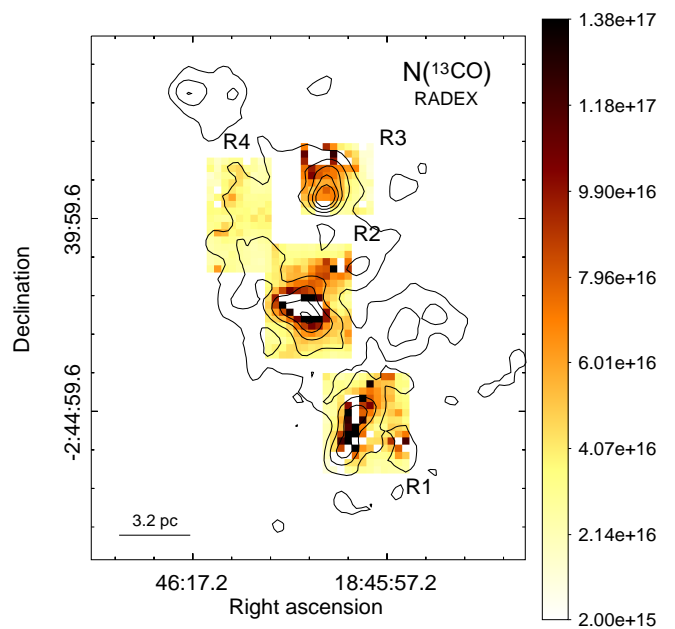}
\caption{\3 column density obtained from the RADEX code towards the analyzed regions. The colourbar is in
units of cm$^{-2}$. Contours of the integrated \8 J=3--2 emission are included for reference.}
\label{radexN13}
\end{figure}

\begin{figure}[h]
\centering
\includegraphics[width=9.3cm]{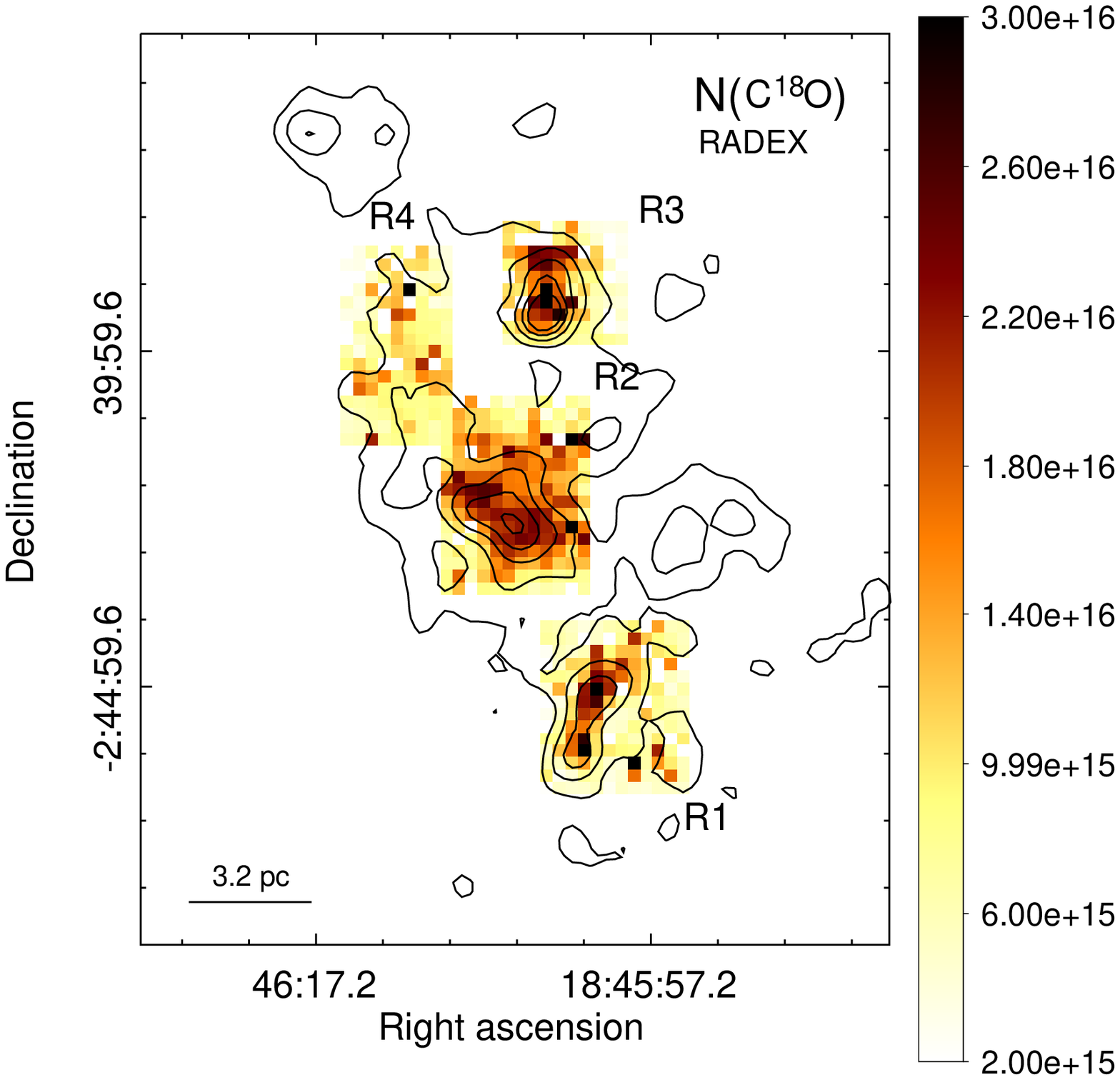}
\caption{\8 column density obtained from the RADEX code towards the analyzed regions. The colourbar is in
units of cm$^{-2}$. Contours of the integrated \8 J=3--2 emission are included for reference.} 
\label{radexN18}
\end{figure}
 
\begin{figure}[h]
\centering
\includegraphics[width=9.3cm]{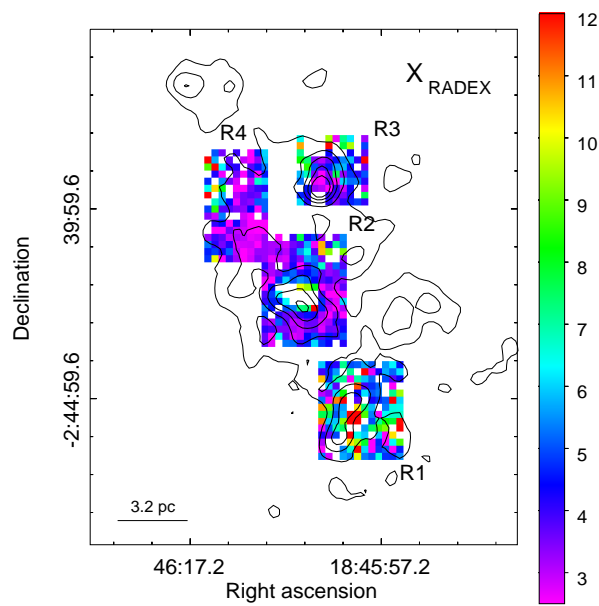}
\caption{\x~abundance ratio obtained from the RADEX code towards the analyzed regions. 
Contours of the integrated \8 J=3--2 emission are included for reference.}
\label{radexX}
\end{figure}

\begin{table}
\caption{Velocity ranges, in \ks, of the selected regions from which second moment and emission peak maps were obtained. }
\label{ranges}
\tiny
\centering
\begin{tabular}{lcccc}
\hline
Region  & \3 2--1  & \3 3--2  & \8 2--1 & \8 3--2  \\
\hline
R1      & 96.8--105.8 & 95.8--106.8 & 97.0--103.0 & 97.0--103.0   \\
R2      & 95.0--103.0 & 95.0--103.0 & 95.0--102.0 & 96.0--102.0  \\
R3     & 93.0--100.0 & 93.0--100.0 & 93.0--100.0 & 93.0--100.0  \\
R4      & 98.0--105.0 & 98.0--105.0 & 99.0--103.5 & 99.0--103.5  \\
\hline
\end{tabular}
\end{table}

The N(\3) and N(\8) obtained from RADEX towards each region are presented in Figures\,\ref{radexN13} and \ref{radexN18},
respectively. There are several pixels for which RADEX does not yield any value (i.e. the code does not converge),
mainly in the \3 case. The reasons for this issue are discussed in the next Section. Figure\,\ref{radexX} displays
the \x~ratio obtained from the N(\3) and N(\8) maps.

\section{Discussion}

\begin{figure}[h]
\centering
\includegraphics[width=9cm]{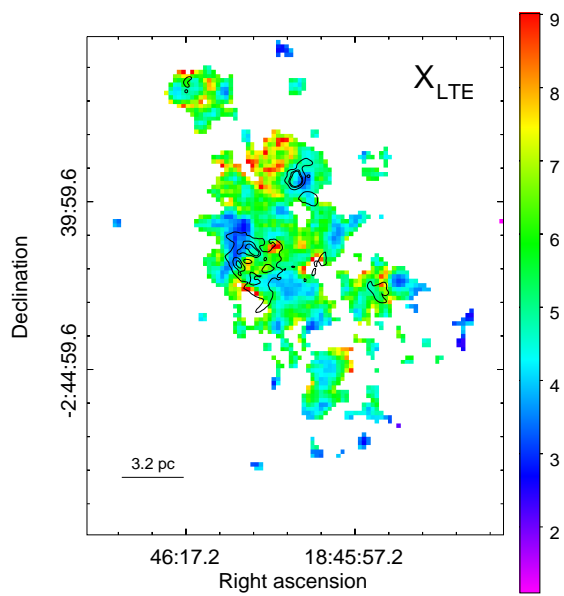}
\caption{Map of the \x~obtained under the LTE assumption with contour of the radio continuum emission with levels of 
15, 40, and 65 mJy beam$^{-1}$. The angular resolution of the radio continuum data is about 5\farcs5. }
\label{xradio}
\end{figure}

Taking into account that \g29~is located at the galactocentric distance of about 4.4 kpc (assuming a distance 
to the source of 6.2 kpc \citep{russ11}), following the 
\citet{wilson94} atomic relations, the \x~ratio in this region should be about 7.
From the LTE results we have an \x~average value of about 5, and as it can be seen in Figure\,\ref{Xlte},
\x~varies from about 1.5 to 10.5 across the studied region. For instance, it can be appreciated that towards the dark cloud 
(marked with 2 in Fig.\,\ref{g29present} (d)) the value of \x~is low, about 3, which is consistent with the fact that 
it should be a dense region in which the UV radiation cannot penetrate to  
selectively photodissociate the \8 isotopes. 
Figure\,\ref{xradio} displays the \x~map obtained from the LTE assumption with contours of the radio continuum emission at 20 cm 
tracing the ionized gas. In general, the higher values of \x~are located mainly at the borders, or close to them, 
of regions with ionized gas, where the far-UV photons are likely escaping and selectively photodissociating the \8 species 
as found in previous works (e.g. \citealt{areal18,kong15,shima14}). This is the case of the northeastern region of the UC \hii~region, where 
there are many pixels with high \x~values (about 8), while towards the southwest, in coincidence with a dark cloud, \x~is low (about 3). 
This spatial distribution suggests that the radiation of the UC \hii~region escapes mainly towards the northeast, photodissociating the gas, 
while towards the west and southwest it is stalling in a dense clump traced by the dust emission at 850 $\mu$m and heating the 
hot core \citep{beuther07}. The same phenomenon may occur in the largest radio continuum feature seen almost at the center in
Fig.\,\ref{xradio}. As mention above, towards the northeast, the UV radiation is encountering a 
very dense component (the dark cloud) which prevents the \8 photodissociation. The same may occur towards the southwest, while 
the far-UV photons likely escape towards the other borders.
Figure\,\ref{plotXvN18} shows that \x~decreases with the increase of the N(\8). Considering that
in general high N(\8) implies high extinction (A$_{\rm V}$), the result displayed in this figure supports that the FUV radiation cannot 
penetrate into 
the inner parts of the clouds (dark clouds) as mention above. Even though we have to be careful with the relation between N(\8) and A$_{\rm V}$ 
because the \8 may be selectively photodissociated in some regions and hence it does not accurately relate to A$_{\rm V}$, 
the tendency displayed in Fig.\,\ref{plotXvN18} is indeed clear.

From the comparison between the maps of the abundance and the integrated line ratios (Figs.\,\ref{Xlte} and \ref{mapIr}),
which compares values that were derived from excitation considerations (LTE assumption) with
values that are direct measurements, it can be appreciated that both parameters present the same behaviour across the region.
In \citet{areal18} we found linear relations between \x~and I$^{13/18}$ that slightly vary among YSOs, \hii, and diffuse \hii~regions.
In this case, a linear tendency can be suggested (see Fig.\,\ref{plotIvX}) but with a large dispersion, which shows 
that very different physical conditions are present in the gaseous component throughout the studied cloud.

The results from RADEX do not cover the whole region as in the LTE case, however we still can obtain valuable information from them. 
Firstly, it can be seen that the column densities values are
larger than those obtained in the LTE assumption, which is something expectable because it is known that LTE column densities of molecular
clouds typically underestimate the true column densities by factors ranging from 1 to 7 \citep{padoan00}.

In the case of \8, RADEX yields results in almost all pixels of the studied regions, and as it can be seen in Fig.\,\ref{radexN18},
the obtained N(\8) towards the selected regions presents similar features as in the N(\8) LTE map. In the \3 case, 
there are several pixels for which RADEX does not yield any value, i.e. pixels where the code does not converge. However, 
it can be seen that pixels with results tend to show similar morphological features as in the N(\3) LTE map. Concerning to the blanked 
pixels, this is not an useless result, 
on the contrary, the non-convergence of the RADEX code can give us important information about the physical conditions in the region.
Analyzing what happens with RADEX in those pixels we found that there are two types of non-convergence: i) the code does not fit the
inputs (T$_{peak}$, $\Delta$v, and T$_{K}$) for one or both lines, and ii) while the inputs of both lines are fitted, the ratio between 
them are not. 

Taking into account that we had assumed that T$_{K} =$ T$_{dust}$, it is possible that the first case of non-convergence
may imply that the gas is not coupled to the dust, mainly in the gaseous component traced mostly by the \3, and thus they 
have different temperatures. Certainly, 
we tested other values of T$_{K}$ and we observed that for values larger than T$_{dust}$, RADEX gives good results. 
The second type of non-convergence
may indicate that the J=2--1 and J=3--2 transitions arise from different components at different temperatures, or even
a more complicate phenomenon: for example it could be that a part of the J=3--2 transition arises from a colder component, 
while another part of the emission arises from a 
hotter one, i.e. a two-component RADEX solution (see \citet{paronN159}, where this was assumed for the J=3--2 transition 
towards an \hii~region in the Large Magellanic Cloud). 
Thus the non-convergence of RADEX shows the complexity of the region, and suggests that the molecular gas is indeed under different physical 
conditions, not only spatially across the cloud, but also along the line of sight.

Considering the above discussion, the \x~values obtained from RADEX (see Fig.\,\ref{radexX}) are in general practically meaningless. 
Only towards the northeast
of R2 and southwest of R4 (where both regions overlap) \x~is quite constant, with values similar to those obtained in the LTE case. 
This region coincides with the dark cloud (marked with 2 in Fig.\,\ref{g29present} (d)), in which the physical conditions may be 
indeed quite constant. 

\subsection{Beam filling factor effects}

It is important to mention that our results are based on the assumption that the beam filling factor of the \3 and \8 
emission is equal to 1, which corresponds to molecular features that completely
fill the telescope beam. This is not completely true because the clouds are usually structured on the subparsec scale implying
beam filling factors less than the unity.
We investigated the influence that, in average, a possible beam dilution effect could have in the obtained X$^{13/18}$.
According to \citet{kim06} we can estimate the beam filling factor as
$\phi=\frac{\theta^2_{\rm source}}{\theta^2_{\rm source}+\theta^2_{\rm beam}}$
where $\theta^2_{\rm source}$ and $\theta^2_{\rm beam}$ are the source and beam sizes, respectively.
The effective beam size of $^{13}$CO and C$^{18}$O J=3--2 data is about 15\s, which corresponds to 0.45~pc at the considered
distance of 6.2~kpc. Using data from {\it Herschel}-PACS 70~$\mu$m (angular resolution $\sim$5\s) we determined that the sizes
of dense cores and/or clumps in the region are expected to exceed 0.7 pc. Thus, the beam filling factor of the used molecular data
is expected to exceed 0.7. Assuming that the \3~emission likely traces more extended molecular components than the \8~emission,
we considered the limit case in which $\phi_{^{13}{\rm CO}}=1$ and $\phi_{\rm C^{18}O}=0.7$. Thus, the obtained \x~could be 
overestimated up to 30\%. 

The  beam dilution effect would be less important in regions of diffuse gas than in those where the gas is clumpy. 
The diffuse gas regions are, in general,  more exposed to the FUV radiation and therefore tends
to have higher values of \x, while the densest/clumpy regions should have lower values of \x. Thus, the beam 
dilution correction, applied mainly to the \8~in clumpy regions, would tend to decrease the \x~value in the densest 
regions, which would highlight even more the behaviour of \x~found accross the cloud.

\section{Summary and concluding remarks}

Using the \2, \3, and \8 J=3--2, and \3, and \8 J=2--1 emission, we studied the column densities and the \3/\8 abundance ratio (\x)
distribution towards the high-mass star-forming region \g29 through LTE and non-LTE methods. 

From the LTE analysis we found an \x~average value of about 5, varying from about 1.5 to 10.5 across the studied region.
This shows that, besides the dependency between \x~and the galactocentric distance, the local physical conditions have 
a great influence on this abundance ratio. In addition, these results show that if one
needs to use a value of \x, it is better to obtain it from the direct molecular observations (if one counts with them)
than from the indirect estimations from the elemental abundances.

Our analysis of the variation in the \x~value across the molecular cloud shows that close to regions of ionized gas
\x~tends to increase, in complete agreement with the selective far-UV \8 dissociation found in other works. Additionally,
the analysis of the spatial distribution between the \x~map, the ionized regions, and the dark 
clouds allows us to suggest in which regions the radiation is stalling in dense gaseous components,
and in which ones it is escaping and selectively photodissociating the \8 species. This roughly maps the 
degree of far-UV irradiation in different molecular components.

The non-LTE analysis based on the RADEX code proves that the region is indeed very complex, showing that the molecular gas may have 
very different physical conditions, not only spatially across the cloud, but also along the line of sight. 
We found that in some cases the non-convergence of RADEX can show when the assumption that gas and dust are thermally coupled fails,
which would define a quite accurate lower limit for the kinetic temperature of the molecular gas.

Finally, it is important to note that this kind of studies may represent a tool to indirect estimate, from
molecular lines observations, the degree of photodissociation in molecular clouds, which is very useful to study
many chemical chains occurring in such interstellar environments.

\section*{Acknowledgments}

We thank the anonymous referee for her/his helpful comments and suggestions.
M.B.A. is a doctoral fellow of CONICET, Argentina. 
S.P. and  M.O. are members of the {\sl Carrera del Investigador Cient\'\i fico} of CONICET, Argentina. 
This work was partially supported by Argentina grants awarded by UBA (UBACyT), CONICET and ANPCYT.

\bibliographystyle{aa}  
\bibliography{ref}
\IfFileExists{\jobname.bbl}{}
{\typeout{}
\typeout{****************************************************}
\typeout{****************************************************}
\typeout{** Please run "bibtex \jobname" to optain}
\typeout{** the bibliography and then re-run LaTeX}
\typeout{** twice to fix the references!}
\typeout{****************************************************}
\typeout{****************************************************}
\typeout{}
}

\label{lastpage}
\end{document}